\documentclass{article}

\pdfoutput=1
\usepackage{fullpage}
\usepackage[utf8]{inputenc}
\usepackage{graphicx}
\usepackage{amssymb}
\usepackage{amsmath}
\usepackage{amsfonts}
\usepackage{tabularx,colortbl}
\usepackage{multirow}
\usepackage{color}
\usepackage[pdftex]{hyperref}
\usepackage[english]{babel}
\usepackage{subfigure}
\usepackage{algorithm}
\usepackage{algorithmic}
\usepackage[sort&compress,numbers]{natbib}
\usepackage{authblk}
\usepackage{appendix}

\definecolor{Orange}{rgb}{1,0.64,0}
\definecolor{lgray}{rgb}{0.92,0.92,0.92}

\makeindex

\begin{document}

\title{Characterization of graphs for protein structure modeling and recognition of solubility}

\author[1]{Lorenzo Livi\thanks{llivi@scs.ryerson.ca}\thanks{Corresponding author}}
\author[2]{Alessandro Giuliani\thanks{alessandro.giuliani@iss.it}}
\author[1]{Alireza Sadeghian\thanks{asadeghi@ryerson.ca}}
\affil[1]{Dept. of Computer Science, Ryerson University, 350 Victoria Street, Toronto, ON M5B 2K3, Canada}
\affil[2]{Dept. of Environment and Health, Istituto Superiore di Sanit\`{a}, Viale Regina Elena 299, 00161 Rome, Italy}
\renewcommand\Authands{, and }
\providecommand{\keywords}[1]{\textbf{\textit{Index terms---}} #1}

\maketitle

\begin{abstract}
This paper deals with the relations among structural, topological, and chemical properties of the E.Coli proteome from the vantage point of the solubility/aggregation propensity of proteins.
Each E.Coli protein is initially represented according to its known folded 3D shape.
This step consists in representing the available E.Coli proteins in terms of graphs.
We first analyze those graphs by considering pure topological characterizations, i.e., by analyzing the mass fractal dimension and the distribution underlying both shortest paths and vertex degrees. Results confirm the general architectural principles of proteins.
Successively, we focus on the statistical properties of a representation of such graphs in terms of vectors composed of several numerical features, which we extracted from their structural representation.
We found that protein size is the main discriminator for the solubility, while however there are other factors that help explaining the solubility degree.
We finally analyze such data through a novel one-class classifier, with the aim of discriminating among very and poorly soluble proteins. Results are encouraging and consolidate the potential of pattern recognition techniques when employed to describe complex biological systems.\\
\end{abstract}

\section{Introduction}
\label{sec:intro}

The delicate balance between solubility and aggregation of protein molecules is a key factor in both protein physiology and disease causation \cite{chiti2002kinetic}. 
Solubility properties allow proteins to work as single molecular systems in cell without exiting from equilibrium in the form of large aggregates as happens with artificial polymers, so being at the very basis of their work as natural and incredibly efficient ``nano-machines'' in the cell.
Highly soluble proteins reach easily their native 3D state (the thermodynamically stable state \cite{ghosh2010cellular,dill2011physical}). On the other hand, individual protein molecules must interact with other proteins in order to generate highly coordinated supra-molecular systems carrying on complex tasks such as DNA duplication and biosynthetic pathways involving ordered reactions sequences. 
From a basic chemico-physical perspective, we can consider solubility as driven by the formation of intra-molecular links (i.e., the molecule folds in order to acquire an energetically favorable configuration in solution \cite{leitner2008solvation}) while aggregation is mainly driven by inter-molecular links.
Both intra- and inter- molecular links have the same chemico-physical nature (hydrogen bonds, van der Waals force etc.).
Besides the physiological aggregation needed for generating supra-molecular systems, proteins can undergo a ``pathological'' aggregation, which is linked to the onset of various diseases \cite{pawar2005prediction}.
The above points illustrate the hardness of the problem to discriminate solubility and aggregation propensities together with its theoretical and applicative importance.

\citet{niwa2009,Niwa05062012} produced a dataset that we consider by far the most unbiased factual basis for facing the solubility/aggregation on a global perspective.
This dataset is made by the entire proteome of E.Coli whose single elements (protein molecules) were assayed as for their relative solubility in cell-free standardized conditions, thus offering for each protein an unbiased measure of solubility.
The more a protein is soluble, the less the fraction that precipitates into insoluble aggregates: this implies that we can consider the Niwa et al. solubility data as a measure of the solubility/aggregation properties of single molecules. This was aptly stated by \citet{Agostini2012237}, approaching the same dataset by means of a sequence-based methodology.
Here we try to go further the purely discriminative task: the quest for a ``global signature'' for protein solubility/aggregation balance was considered as a vantage point for a testing of different network-based representations of protein 3D structures, going from pure topology to labeled graphs enriched with a chemico-physical description of amino acid residues.
The emerging correlation structure among different graph invariants confirmed some established architectural principles of proteins as their fractal and modular nature \cite{di2012proteins,banerji2011fractal,PhysRevE.71.011912,liang2001proteins,tejera2009fractal,tasdighian2013modules,doi:10.1021/cr3002356,clune2013evolutionary}.
On the other hand, the strict dependence of protein structural, topological, and solubility properties on their size confirmed a general principle of nanomaterials \cite{klabunde2001nanoscale,sanchez1999gold}, whose collocation in-between molecular and bulk material worlds makes their physico-chemical behavior strictly dependent on size. 
Proteins and supra-molecular aggregates live at the nanoscale and the recognition of a strict size dependence of their properties is an evidence for the possible cross-fertilization of protein and nanomaterials science.

Providing descriptive characterizations of complex systems has a long history in the field of (complex) dynamical systems and chaos theory \cite{wackerbauer1994comparative,nakayama1994dynamical,giuliani2001complexity}.
Describing (complex) systems by means of graphs is ubiquitous in modern science and engineering disciplines \cite{newman2010networks,csermely2013structure,costa2007characterization,boccaletti+latora+moreno+chavez+hwang2006,porfiri2008synchronization,donner2011,4196541,bullmore2009complex,dehmer2013quantitative,giuliani2014network}.
In fact, graphs offer a sound mathematical framework to describe the relations/causality among the interacting elements of the system under analysis.
Characterizing a graph by numerical values (e.g., numerical features/descriptors) is usually based on suitable graph-theoretic results, adaptation of information-theoretic concepts, or by exploiting interpretations in terms of dynamical processes (such as diffusion, percolation, and state transition).
The profound multi-disciplinary character of the topic produced a number of interpretations of similar concepts, although by exploiting several different techniques.
It is possible to cite approaches involving fractal analysis, percolation, and (anomalous) diffusion \cite{ben2000diffusion,havlin2010,song2005self,daqing2011dimension,song2007calculate}.
It is worth citing also the more recently-proposed graph characterization of \citet{escolano2012heat} (see also \cite{Bai20141172} for related material), called flow complexity (based on thermodynamic depth \cite{lloyd1988complexity}).
Random walk based approaches are also highly popular. For instance, random walks are used to model the interaction of ``information waves'' in a graph \cite{mirshahvalad2014dynamics}, which is useful to describe at the same time the spread and the interaction of information over time.
Analogously, the concepts of hitting (average probability that two random walkers are in the same state at the same time) and commute time (average return time to the initial state) in random walks have been used by \citet{Qiu20072874} to characterize graphs for pattern recognition purpose.
Other approaches include computation of Ihara coefficients and Laplacian spectrum \cite{5648360,seriation+gradis_lncs_2012}.
Finally, various interpretations of the concept of network entropy are studied \cite{Dehmer201157}, such as entropy of continuous time quantum walks \cite{mulken2011continuous,PhysRevE.80.045102}, network ensemble entropy \cite{bianconi2009entropy,PhysRevE.80.045102}, network transfer entropy \cite{PhysRevE.87.052814}, von Neumann (quantum) entropy \cite{Han20121958,ye2014approximate}, and fuzzy entropy \cite{Livi_ga_2013}.

In this paper, we exploit several graph-based techniques to model protein structures and recognize the important solubility property of the E.Coli proteins from \citet{niwa2009}. The analysis is based on a dataset previously elaborated by us \cite{ecoli_graph}.
Initially, we provide a pure topological characterization of the graph structures by exploiting fractal analysis techniques and by considering the underlying distribution of both shortest paths and degrees.
Successively we elaborate a new, vector-based, representation of the graphs. For this purpose, we extract 15 different features (graph characteristics). Such a new protein representation is hence analyzed in this paper with the aim of understanding the statistical relations among the structure and the solubility degree of the E.Coli proteins.
First, we provide a factor analysis of the dataset of 15 features here elaborated. Results confirm the well-known fact that protein size is the most important factor. At the same time, we complement the linear correlation structure by analyzing also the non-linear relations via the estimation of the mutual information.
Finally, we face the problem of discriminating among very soluble and poorly soluble proteins. Notably, we deal with the problem in the one-class classification setting \cite{eocc}.
From a bioinformatics viewpoint, the paper demonstrates how graphs are powerful modeling and computational frameworks, allowing for a hybrid style of analysis linking purely statistical and content-related features of complex systems.

The remainder paper is structured as follows. Section \ref{sec:ecoli_graph} describes the protein representation in terms of labeled graphs. In Section \ref{sec:complexity_descriptors} we introduce the features (characteristics) that we have extracted to represent each graph as real-valued vector. In Section \ref{sec:experiments} we discuss the experimental results. Section \ref{sec:conclusions} concludes the paper.
Finally, Appendix \ref{sec:graph_descriptors} provides the essential details of the considered graph characteristics.

\section{Graph Representation of E.Coli Proteins}
\label{sec:ecoli_graph}

In our previous work \cite{ecoli_graph}, we elaborated the E.Coli data of \citet{niwa2009,Niwa05062012} by constructing a graph representation. We were able to retrieve the 3D structure from the Protein Data Bank (PDB) \cite{pdb} of 454 proteins among the 3173 originally provided by Niwa et al. Let us refer to this dataset of graphs (\cite{ecoli_graph}) as DS-G-454.
The contact graph of a protein is constructed by mapping each amino acid residue to a vertex of the resulting graph.
An edge is added among two amino acids if the Euclidean distance between the two centers of mass in the 3D space is between 4 and 8 \AA{}, so to filter out trivial contacts among neighboring residues along the sequence.
Vertices are equipped with suitable attributes (labels) that are defined as the three principal components (PCs) derived from the analysis of several chemico-physical properties of amino acids \cite{apdbase}. Edges are equipped with the Euclidean distance among residues.

\begin{figure}[ht!]
 \centering
 \includegraphics[bb=0 0 410 302,scale=0.6,keepaspectratio=true]{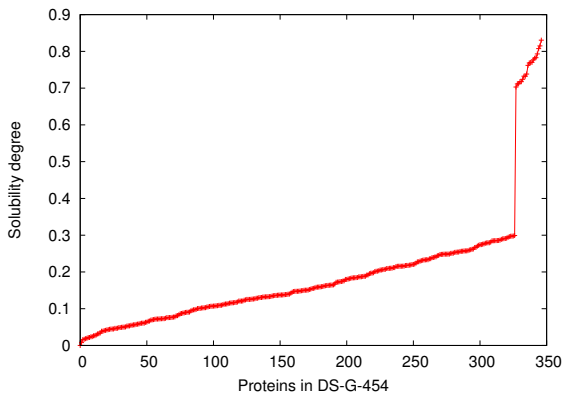}
 \caption{Solubility of the 454 proteins in DS-G-454 sorted in increasing order.}
 \label{fig:solubility_ds-g-454}
\end{figure}

Originally, such proteins are associated to a continuous solubility degree, which, after a straightforward normalization, can be considered as a number in $[0, 1]$; 0 indicating the lowest solubility, 1 the highest in the original data.
Fig. \ref{fig:solubility_ds-g-454} shows the solubility degree of the proteins arranged in increasing order. Although the solubility is a chemical property defined within a continuous domain, it is possible to observe that in DS-G-454 there is an evident demarcation among two classes of proteins: those that are very soluble and those are highly insoluble. In particular, we have 77 highly-soluble proteins and 377 with low solubility propensity.
This fact allows us to make the reasonable assumption that the proteins in DS-G-454 are members of two distinct classes: soluble and insoluble. This diversification will be studied in the following with the aim of providing a statistical and structural characterization of the E.Coli proteins with respect to the solubility property.

\section{The Considered Graph Features}
\label{sec:complexity_descriptors}

This section briefly introduces the considered features (graph characteristics) that we extracted from the graphs in DS-G-454.
Technical details are reported in Appendix \ref{sec:graph_descriptors}.

Each graph in DS-G-454 is described as a real-valued vector of 15 components, forming hence a dataset that we refer to as DS-C-454. The features are: number of vertices (V), number of edges (E), number of chains in the molecule (C), radius of gyration (RG), porosity (P), modularity of the graph (M), average closeness centrality (ACC), average degree centrality (ADC), and average clustering coefficient (ACL), energy (EN) and Laplacian energy (LEN) of the graph, heat trace (HT) and heat content invariant (HCI), the ambiguity (A), and finally the entropy of the stationary distribution (H).

All such features taken together cover different aspects of the proteins when represented as graphs.
Of course, number of vertices/edges (i.e., number of residues and related contacts) accounts for the most basic characterization of proteins: their size. However, as we will see in the following, protein size is one of the most important factors when considering the solubility/aggregation propensity.
Number of chains is an important protein characteristic, since in fact the different chains of a protein fold separately and only at end of the process they bind together to function.
RG and P are two features describing the ``shape'' of a protein. In fact, RG is a measure describing the distribution of residues around the center of mass of the whole molecule, while P describes the compactness of the structure (it is a measure of the empty space within the molecule boundaries).
M, ACC, ADC, and CL are either local and global structural descriptors that characterize the topology of the graphs (in terms of paths and local structure). In particular, M has been computed by using the \textit{k}-means algorithm (best \textit{k} strategy is used, with $2\leq k \leq 15$).
EN, LEN, HT, and HCI are more sophisticated features that are extracted from the spectrum of the matrix representation of the graphs (adjacency and Laplacian matrices). HT is computed for $t=5$ (as a compromise among local and global description) while HCI for $m=1$ (we consider the first coefficient only).
Those features have a long history in chemistry and the physical sciences \cite{trinajstiac1983chemical,gutman2006laplacian,lervik2010heat}, since they have been used to model real processes such as diffusion and percolation \cite{ben2000diffusion}.
Finally, a measure of irregularity of the graph (how much the structure of a graph differs from a regular graph) is provided by A, and H synthetically describes the unpredictability of a Markovian random walk on a graph -- a complete graph induces a transition matrix following the uniform distribution; the most uncertain distribution.
The unpredictability is quantified as the 2-order R\'{e}nyi entropy of the stationary distribution.

\section{Protein Structure Analysis and Recognition of Solubility}
\label{sec:experiments}

In this section, we study different aspects of the 454 E.Coli proteins represented according to both DS-G-454 and DS-C-454.
First, we study two important topological properties of the E.Coli graphs in DS-G-454: (i) the fractal dimension of the embedded graphs and (ii) the distributions of both shortest paths and vertex degrees (Sec. \ref{sec:geometry}).
In Sec. \ref{sec:dependence}, we study the statistical relations among the elaborated 15 graph features (data in DS-C-454). First we look for a more interpretable linear correlation structure by means of a factor analysis (\ref{sec:factors}). Then, we move to a non-linear setting by estimating the mutual information among the variables (\ref{sec:mi}).
Finally, in Sec. \ref{sec:recognition} we analyze DS-C-454 in terms of recognition/discrimination capability of soluble/insoluble proteins.

\subsection{Topological Structure of E.Coli Proteins}
\label{sec:geometry}

\subsubsection{Fractal Dimension of Proteins from Radius of Gyration}
\label{sec:mfd}

\begin{figure*}[ht!]
\centering

\subfigure[log-log plot for soluble proteins.]{
\includegraphics[viewport=0 0 346 241,scale=0.6,keepaspectratio=true]{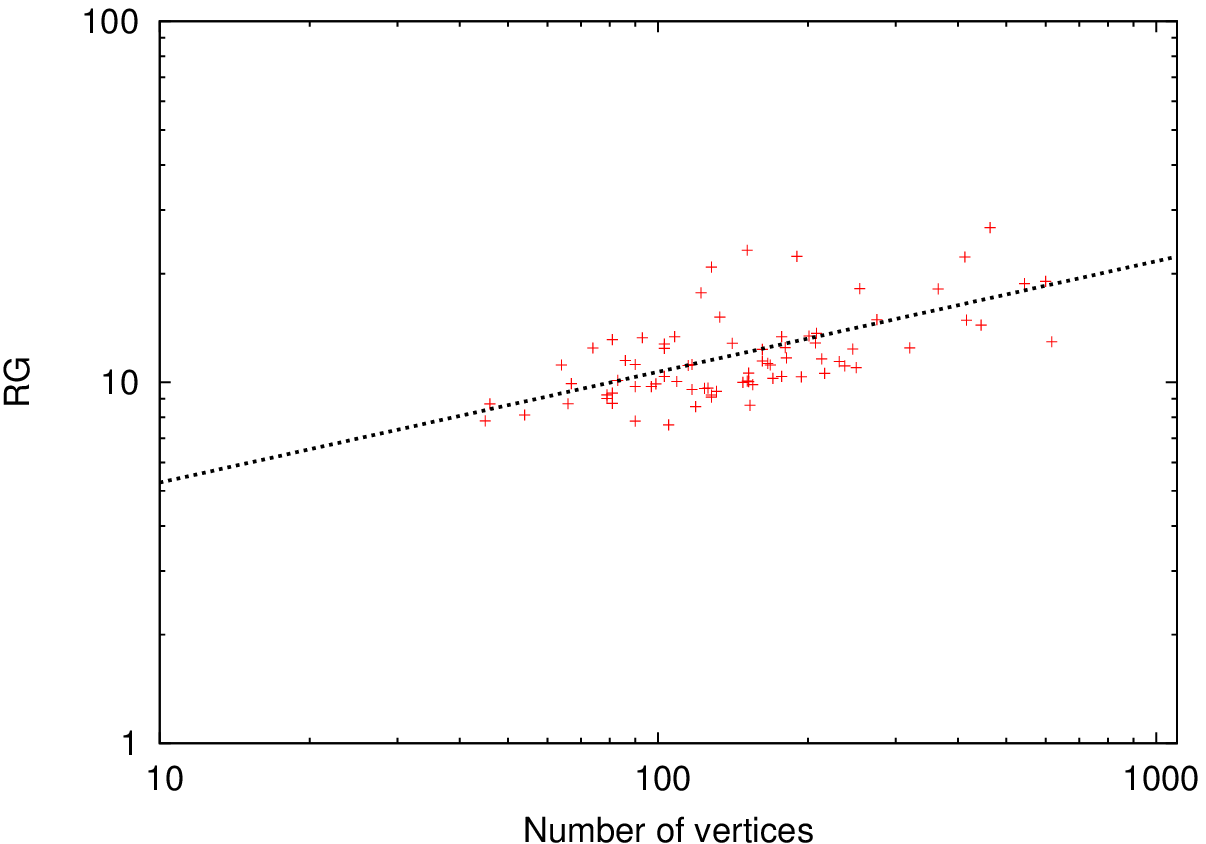}
\label{fig:mfd_sol}}
~
\subfigure[log-log plot for insoluble proteins.]{
\includegraphics[viewport=0 0 346 241,scale=0.6,keepaspectratio=true]{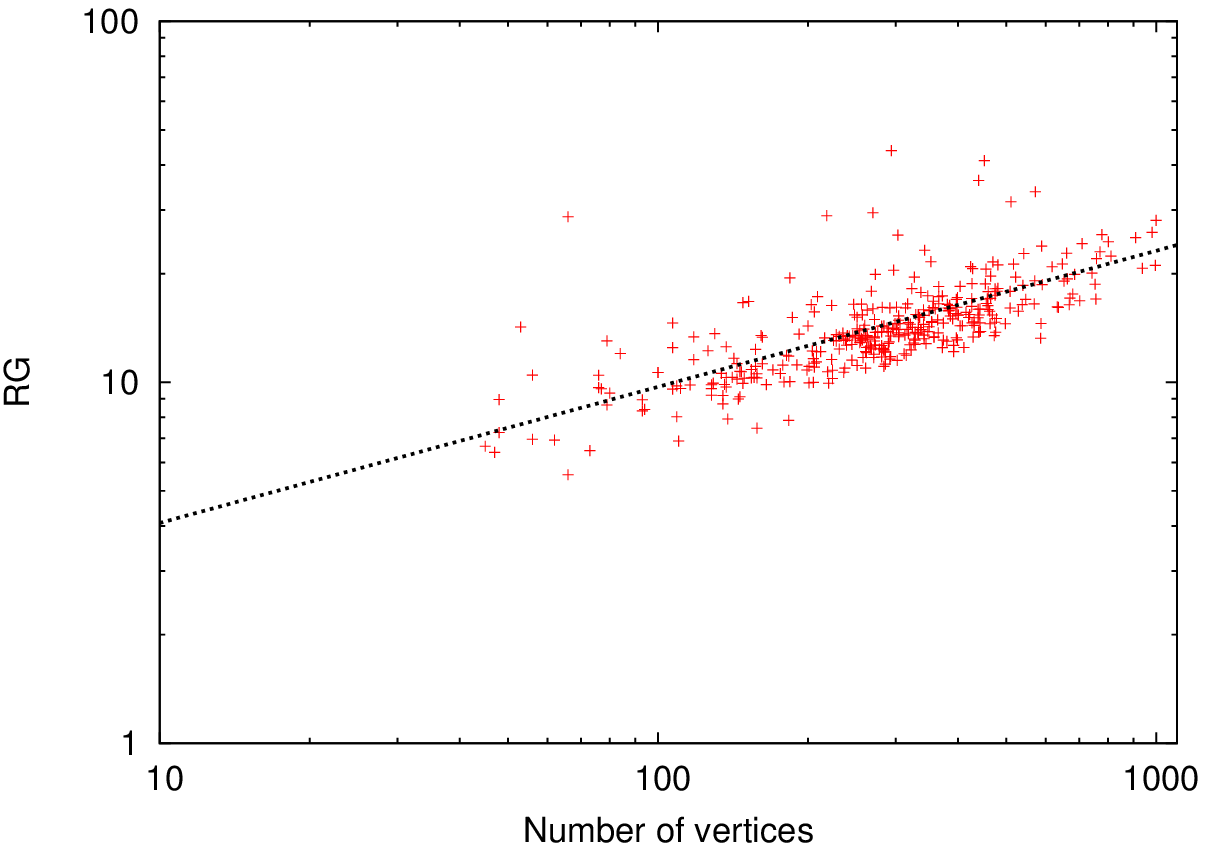}
\label{fig:mfd_ins}}
~
\subfigure[log-log plot for all proteins in DS-G-454.]{
\includegraphics[viewport=0 0 346 241,scale=0.6,keepaspectratio=true]{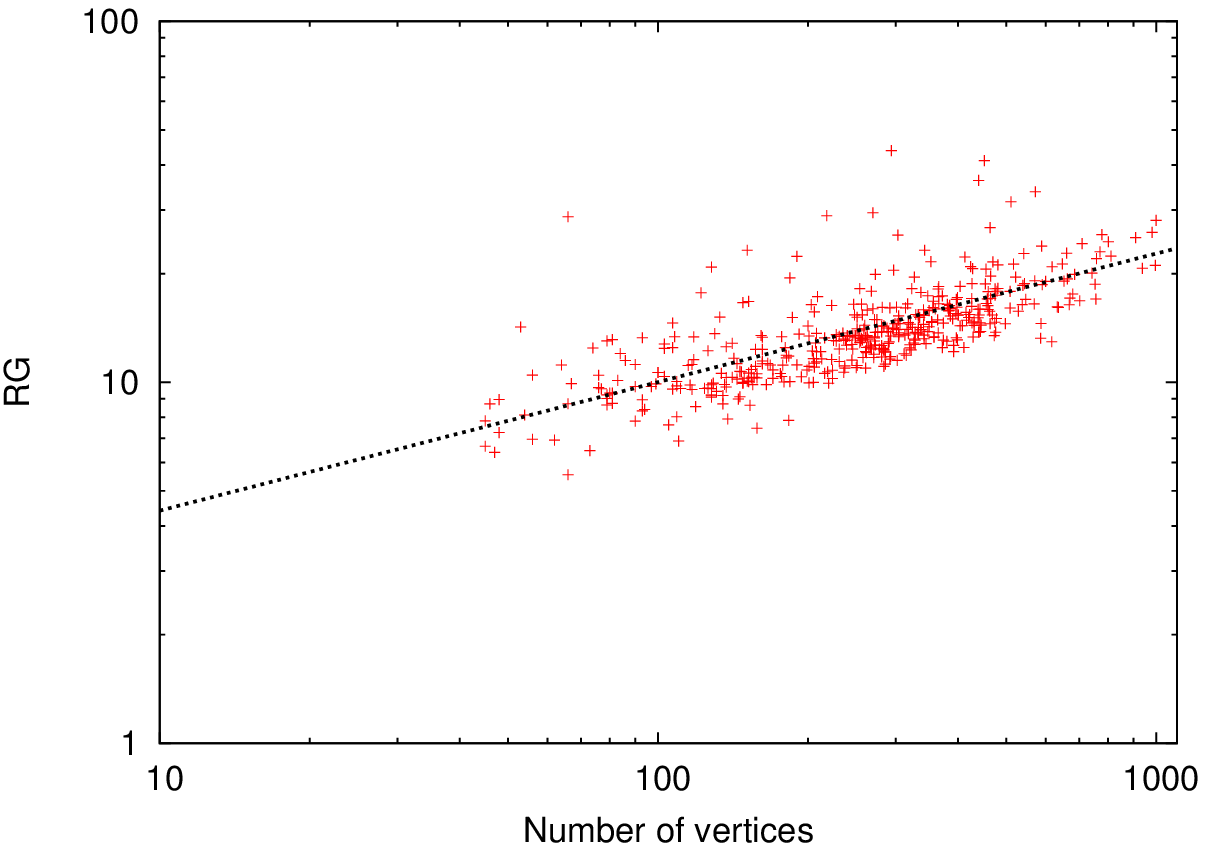}
\label{fig:mfd_all}}

\caption{MFD calculations from RG. In the figures, we show the log-log plots from which we have calculated the MFD of soluble (\ref{fig:mfd_sol}), insoluble (\ref{fig:mfd_ins}), and all proteins taken together (\ref{fig:mfd_all}).}
\label{fig:mfd_rg}
\end{figure*}

Here we discuss the mass fractal dimension (MFD) determined from the scaling among the mass of residues and RG of the molecule \cite{di2012proteins,PhysRevE.71.011912}.
MFD can be computed for a single protein or for a collection of proteins. In the second case, it is possible to derive the MDF, $D$, by studying the scaling among RG, $R_{G}$, and the mass/length, $M$, of the polymer: $R_{G}\sim M^{1/D}$.
Fig. \ref{fig:mfd_rg} shows the log-log plots that we used to determine the MFDs.
We performed the experiments by either separating soluble and insoluble proteins, and also by considering at the same time all proteins in DS-G-454.
MFD of soluble proteins is $\simeq 3.2$; for the insoluble is $\simeq 2.6$; considering all proteins we have $\simeq 2.8$.
At first, one might conclude that MDF offers an interesting mechanism for discriminating soluble and insoluble proteins.
However, the low coefficients of determination ($\simeq 0.4$ for soluble; $\simeq 0.52$ for insoluble) suggest us that such a feature cannot be considered as a reliable class discriminator -- intra-class agreement on such a measure is not strong. In fact, although RG conveys important information (see Sec. \ref{sec:factors} and \ref{sec:mi}), it does not offer a striking discrimination rule for the solubility.

We computed the MFD also for each single protein (data not shown) by analyzing the scaling $M\sim R^{D}$, considering hence the mass, $M$, of the atoms falling within concentric spheres of characteristic length $R$. Spheres are centered at the center of mass of the molecule.
In average, we obtained a MFD of $\simeq 2.5$, which is fairly smaller than the one obtained with the RG. This difference could be explained by the not so strong fitting precision observed in the scaling among vertices and RG.
Nonetheless, considering the average size of the proteins at hand, this value is in agreement with the calculations reported in the literature \cite{di2012proteins,PhysRevE.71.011912}.

\subsubsection{Distance and Degree Distributions}
\label{sec:distances}

\begin{figure*}[ht!]
\centering

\subfigure[Scaling of ACC with the number of vertices.]{
\includegraphics[viewport=0 0 340 243,scale=0.6,keepaspectratio=true]{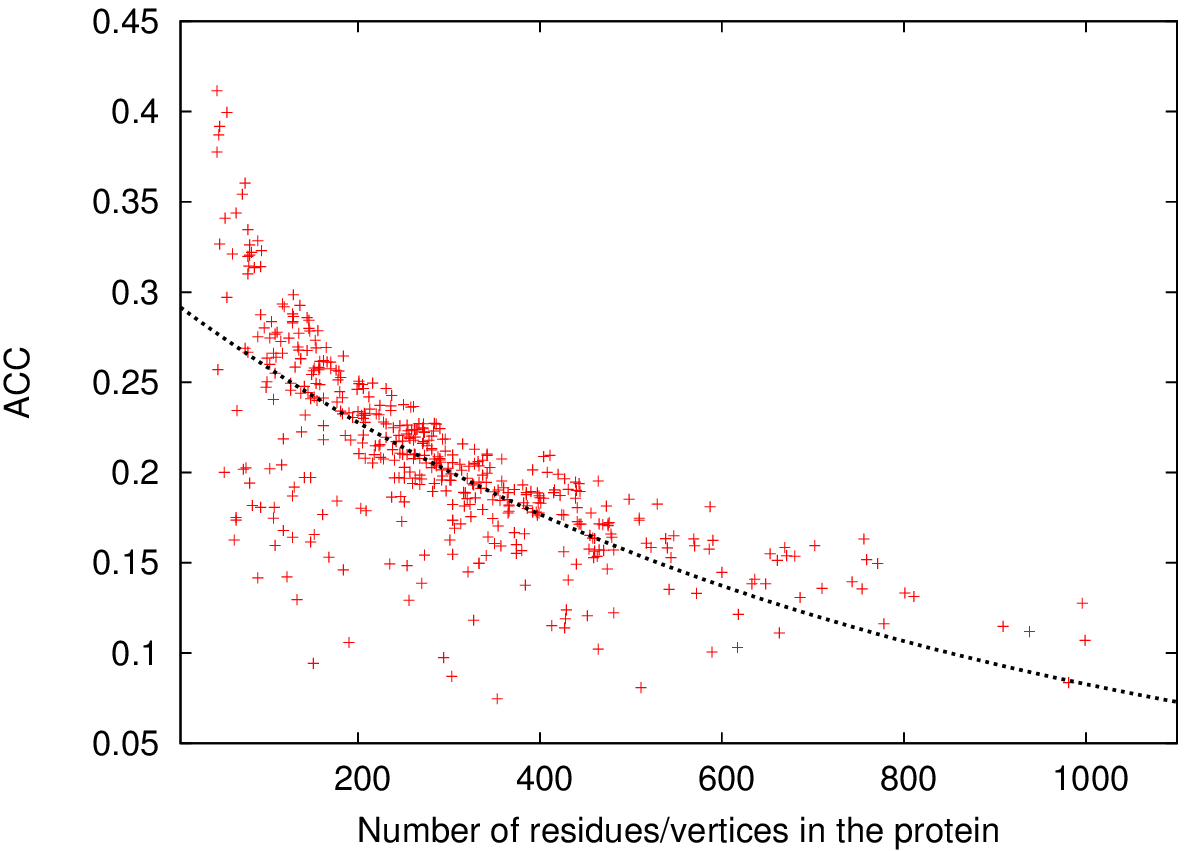}
\label{fig:closeness_centr}}
~
\subfigure[Degree distribution.]{
\includegraphics[viewport=0 0 345 243,scale=0.6,keepaspectratio=true]{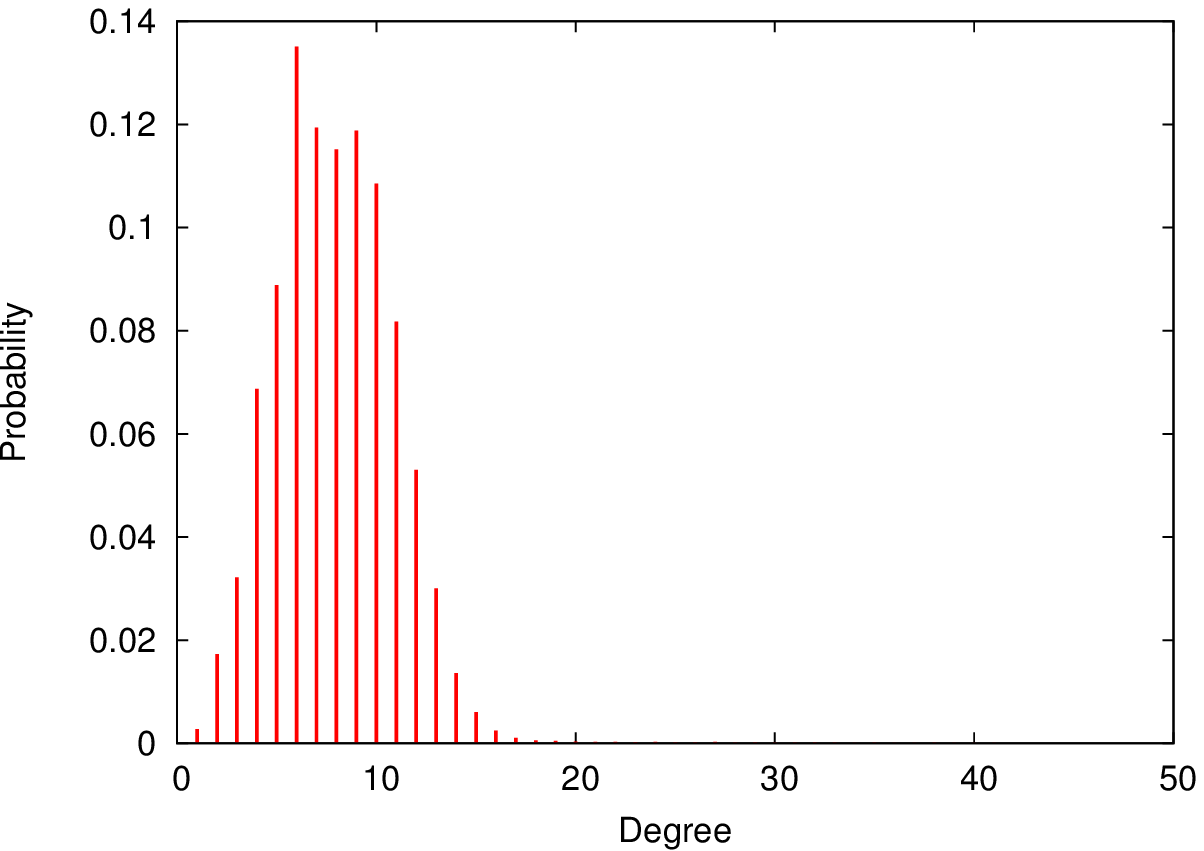}
\label{fig:degr_distr}}

\caption{Distances (\ref{fig:closeness_centr}) and degrees (\ref{fig:degr_distr}) of the proteins graphs. All 454 proteins are analyzed together.}
\label{fig:distances_degrees}
\end{figure*}

As already observed by \citet{tasdighian2013modules}, proteins exhibit a typical topology that is at the same time highly modular, fractal, and usually hierarchical. This results in networks that are also optimized in terms of shortest paths, while however the characteristic path length of proteins does not scale with high agreement as a small-world topology \cite{bartoli2007effect,doi:10.1021/cr3002356,PhysRevE.71.011912}.
As a demonstration of this fact for our dataset, DS-G-454, in Fig. \ref{fig:closeness_centr} we show the scaling between the number of vertices and the related ACC.
As it is possible to observe, the scaling is not very consistent with an exponential decay -- please note that an analogue result can be obtained by considering the average shortest path, but observing an increasing logarithmic-like trend.
This is in agreement with the fact that the topology is not entirely small-world.

Proteins topology should not be considered scale-free as well, since in fact the presence of hubs is limited \cite{bagler2005network,bode2007network,yan2014construction}. This is due to the fact that the physical arrangement of neighbor residues is limited by steric effects. As a demonstration of this fact, in Fig. \ref{fig:degr_distr} we show the sample degree distribution, which is clearly not consistent with an inverse power-law.

\subsection{Statistical Analysis of Features}
\label{sec:dependence}

\subsubsection{Factor Analysis}
\label{sec:factors}

The loadings of the first five factors (components extracted by PCA on the correlation matrix) with the original descriptors are reported in Tab. \ref{tab:factors}. The variables relevant for component evaluation (higher loading in module) are in bold. The extracted components are consistent with the correlation structure made evident in Ref. \cite{di2012proteins}.
By far, the most important order parameter shaping the protein universe is their relative size, Factor 1, which is heavily correlated with explicit size features, such as V, E, and RG.
Moreover, as demonstrated in Sec. \ref{sec:distances}, ACC scales negatively with the size, i.e., larger proteins have a lower average closeness centrality factor.
It is worth pointing out the strong (linear) relation of some size-related variables with descriptors such as EN, LEN, HT, and HCI. Although these features obviously depend also on the number of vertices/edges, they convey higher level information that characterize also the global arrangement of the network.
This result confirms us that all proteins share common mesoscopic architectural principles.
The consequences of these architectural principles in terms of protein properties are in turn modulated by the effect of size as expected with nanoscale objects, as protein molecules are \cite{klabunde2001nanoscale}.

It is interesting to note that the number of chains of a protein (C) is independent of the size component (Factor 1), while it needs a dedicated component (Factor 4). On the same component, it is worth mentioning the weaker contribution of CL, pointing out the fact that the local clustering coefficient is lower in multiple-chains molecules.

Factor 2 is the most relevant order parameter after the predominant size component (result in accordance with Ref. \citet{di2012proteins}). In particular, it points to peculiar topological properties of proteins (ADC, CL). It is worth noting the fact that A is almost equally loaded on the size component and the second factor. Notably, A decreases with the network size and it increases as the network becomes more irregular (in terms of ADC and CL).

Again in accordance with the general results presented in \cite{di2012proteins}, the third factor emerges as a ``global shape'' factor describing the arrangements of the amino acids in the molecule. In fact, P is almost entirely loaded in this component.
It is worth noting that the local topological properties (Factor 2) are independent from the global architecture (Factor 3).
We already commented Factor 4 (number of chains). Factor 5 allows a less clear interpretation, while however it is worth noting that H scales almost identically here and with the size component (Factor 1).
\begin{table*}[tph!]\footnotesize
\caption{Loadings of the first five factors of DS-C-454.}
\begin{center}
\begin{tabular}{|c|c|c|c|c|c|}
\hline
\rowcolor{lgray} & \textbf{Factor 1} & \textbf{Factor 2} & \textbf{Factor 3} & \textbf{Factor 4} & \textbf{Factor 5} \\
\hline
\cellcolor{lgray}\textbf{V} & \textbf{0.96256} & 0.18522 & -0.06749 & -0.06373 & 0.04916 \\
\hline
\cellcolor{lgray}\textbf{E} & \textbf{0.95995} & 0.03505 & -0.17507 & 0.02587 & 0.07179 \\
\hline
\cellcolor{lgray}\textbf{C}  & 0.18263 & -0.18991 & 0.15510 & \textbf{0.85074} & 0.31673 \\
\hline
\cellcolor{lgray}\textbf{RG} & \textbf{0.68604} & 0.29540 & 0.04067 & 0.01040 & 0.17071 \\
\hline
\cellcolor{lgray}\textbf{P} & 0.08963 & -0.02784 & \textbf{0.84408} & -0.16946 & 0.33229 \\
\hline
\cellcolor{lgray}\textbf{M} & \textbf{0.65718} & -0.12567 & 0.27276 & -0.07251 & -0.07648 \\
\hline
\cellcolor{lgray}\textbf{ACC} & \textbf{-0.81237} & 0.15999 & -0.26565 & 0.07573 & 0.07720 \\
\hline
\cellcolor{lgray}\textbf{ADC} & \textbf{0.58607} & \textbf{-0.59391} & -0.24486 & 0.14777 & 0.02664 \\
\hline
\cellcolor{lgray}\textbf{CL} & -0.17661 & \textbf{-0.56592} & -0.18399 & \textbf{-0.46088} & \textbf{0.49981} \\
\hline
\cellcolor{lgray}\textbf{EN} & \textbf{0.96986} & 0.13734 & -0.09073 & -0.05360 & 0.05268 \\
\hline
\cellcolor{lgray}\textbf{LEN} & \textbf{0.95349} & 0.07679 & -0.17896 & 0.01967 & 0.12255 \\
\hline
\cellcolor{lgray}\textbf{HT} & \textbf{0.92012} & 0.31522 & -0.02426 & -0.07567 & 0.06625 \\
\hline
\cellcolor{lgray}\textbf{HCI} & \textbf{0.95737} & 0.19831 & -0.05844 & -0.05293 & 0.05536 \\
\hline
\cellcolor{lgray}\textbf{A} & \textbf{-0.70970} & \textbf{0.63644} & 0.00745 & 0.01040 & 0.14565 \\
\hline
\cellcolor{lgray}\textbf{H} & \textbf{0.57668} & -0.26247 & 0.27621 & -0.01511 & \textbf{-0.54429} \\
\hline
\end{tabular}
\label{tab:factors}
\end{center}
\end{table*}

\subsubsection{Non-linear Correlations via Mutual Information}
\label{sec:mi}

The linear correlation structure discussed in Sec. \ref{sec:factors} confirms the distinguishing factor of size in proteins.
In fact, the first factor is highly characterized by either common size descriptive attributes (such as V, E, and RG) and other related characteristics, such as EN, LEN, HT, and HCI (although these features account also for more sophisticated characteristics, such as the global network structure and the flow of information).

Here we elaborate over those results by analyzing the pairwise non-linear correlations. We compute the non-linear correlation among two features/characteristics via mutual information estimation \cite{Moon_MI__2002}. Results are provided in Tab. \ref{tab:mi}.
Overall, the correlation structure is very similar to the one discussed in (\ref{sec:factors}). In fact, V, E, RG, EN, LEN, HT, are HCI are highly correlated to each other. In addition, P exhibits weak relations with respect to all other features, which justifies the fact that in the factor analysis this feature was loaded alone in a dedicated component (Factor 3).
Individual correlations with respect to the solubility degree (SOL) show that there is no predominant strong non-linear relation. However, results in bold seems to share more information with this target property.
\begin{table*}[tph!]\footnotesize
\caption{Pairwise mutual information estimated from the features in DS-C-454. First row shows the individual contributions with respect to the solubility degree, SOL. The remaining entries form a symmetric matrix conveying the pairwise non-linear statistical dependence calculated via the mutual information estimation.}
\begin{center}
\begin{tabular}{|c|c|c|c|c|c|c|c|c|c|c|c|}
\hline
\rowcolor{lgray}  & \textbf{V} & \textbf{E} & \textbf{C} & \textbf{RG} & \textbf{P} & \textbf{M} & \textbf{ACC} & \textbf{ADC} & \textbf{CL} & \textbf{EN} & \textbf{LEN}\\
\hline
\cellcolor{lgray}\textbf{SOL} & \textbf{0.4404} & \textbf{0.4389} & 0.1840 & 0.3521 & 0.1492 & 0.3557 & \textbf{0.4375} & 0.3767 & 0.2549 & \textbf{0.4388} & \textbf{0.4450} \\
\hline
\cellcolor{lgray}\textbf{V} & 1.0000 & 0.9917 & 0.3349 & 0.8235 & 0.2508 & 0.7409 & 0.9366 & 0.6675 & 0.4715 & 0.9987 & 0.9925 \\
\cellcolor{lgray}\textbf{E} & 0.9917 & 1.0000 & 0.3496 & 0.8067 & 0.2675 & 0.7236 & 0.9235 & 0.7411 & 0.4679 & 0.9961 & 0.9973 \\
\cellcolor{lgray}\textbf{C} & 0.3349 & 0.3496 & 1.0000 & 0.3412 & 0.2210 & 0.3296 & 0.3162 & 0.3114 & 0.2853 & 0.3354 & 0.3598 \\
\cellcolor{lgray}\textbf{RG} & 0.8235 & 0.8067 & 0.3412 & 1.0000 & 0.3210 & 0.7271 & 0.7795 & 0.4763 & 0.5015 & 0.8135 & 0.8153 \\
\cellcolor{lgray}\textbf{P} & 0.2508 & 0.2675 & 0.2210 & 0.3210 & 1.0000 & 0.2604 & 0.3126 & 0.2296 & 0.2816 & 0.2530 & 0.2615 \\
\cellcolor{lgray}\textbf{M} & 0.7409 & 0.7236 & 0.3296 & 0.7271 & 0.2604 & 1.0000 & 0.7703 & 0.5241 & 0.4449 & 0.7347 & 0.7290 \\
\cellcolor{lgray}\textbf{ACC} & 0.9366 & 0.9235 & 0.3162 & 0.7795 & 0.3126 & 0.7703 & 1.0000 & 0.6657 & 0.5270 & 0.9335 & 0.9261 \\
\cellcolor{lgray}\textbf{ADC} & 0.6675 & 0.7411 & 0.3114 & 0.4763 & 0.2296 & 0.5241 & 0.6657 & 1.0000 & 0.5145 & 0.6996 & 0.7276 \\
\cellcolor{lgray}\textbf{CL} & 0.4715 & 0.4679 & 0.2853 & 0.5015 & 0.2816 & 0.4449 & 0.5270 & 0.5145 & 1.0000 & 0.4675 & 0.4751 \\
\cellcolor{lgray}\textbf{EN} & 0.9987 & 0.9961 & 0.3354 & 0.8135 & 0.2530 & 0.7347 & 0.9335 & 0.6996 & 0.4675 & 1.0000 & 0.9952 \\
\cellcolor{lgray}\textbf{LEN} & 0.9925 & 0.9973 & 0.3598 & 0.8153 & 0.2615 & 0.7290 & 0.9261 & 0.7276 & 0.4751 & 0.9952 & 1.0000 \\
\hline
\end{tabular}
\label{tab:mi}
\end{center}
\end{table*}
\begin{table*}[tph!]\footnotesize
\caption{Continue of Tab. \ref{tab:mi}.}
\begin{center}
\begin{tabular}{|c|c|c|c|c|}
\hline
\rowcolor{lgray} & \textbf{HT} & \textbf{HCI} & \textbf{A} & \textbf{H} \\
\hline
\cellcolor{lgray}\textbf{SOL} & \textbf{0.4341} & \textbf{0.4415} & \textbf{0.4078} & 0.3898 \\
\hline
\cellcolor{lgray}\textbf{HT} & 1.0000 & 0.9871 & 0.6491 & 0.6953 \\
\cellcolor{lgray}\textbf{HCI} & 0.9871 & 1.0000 & 0.7227 & 0.7273 \\
\cellcolor{lgray}\textbf{A} & 0.6491 & 0.7227 & 1.0000 & 0.7638 \\
\cellcolor{lgray}\textbf{H} & 0.6953 & 0.7273 & 0.7638 & 1.0000 \\
\hline
\end{tabular}
\label{tab:mi_2}
\end{center}
\end{table*}

\subsection{Discrimination of Solubility Classes}
\label{sec:recognition}

In this section, we focus on DS-C-454 with the aim of providing a robust discrimination (i.e., classification) rule for soluble and insoluble proteins.
As stressed before, we faced this objective multiple times in previous studies \cite{grapsec_ijcnn_2013,ecoli_graph}, but always considering a multi-class approach, i.e., by modeling explicitly soluble and insoluble classes.
In this paper, we depart from this approach by casting the problem in terms of one-class classification.
To this end, we will make use of a one-class classifier that we recently proposed \cite{eocc}. In the following, we refer to this classifier as EOCC.
We consider the very soluble proteins as the ``targets'', while those that are insoluble are considered as non-target.
This choice is justified by the fact that all proteins for which it is available the 3D structure must be solubilized.
Therefore, a low level of solubility, measured according to the \citet{niwa2009} conditions, can be understood as an anomaly; in practice, those proteins with lower solubility levels are stabilized in the E.Coli organism by the so-called chaperones.
The aim of the test is to provide consistent acceptance/rejection decisions considering the fact that proteins are originally described by a continuous variable: the solubility degree.
EOCC provides both hard and soft decisions, which perfectly suits this scenario. We will see that the soft decisions convey reasonable information with respect to the original solubility degree.

\subsubsection{Data Preprocessing and Organization}

\begin{figure*}[ht!]
\centering

\subfigure[PC1--PC2.]{
\includegraphics[bb=0 0 343 245,scale=0.65,keepaspectratio=true]{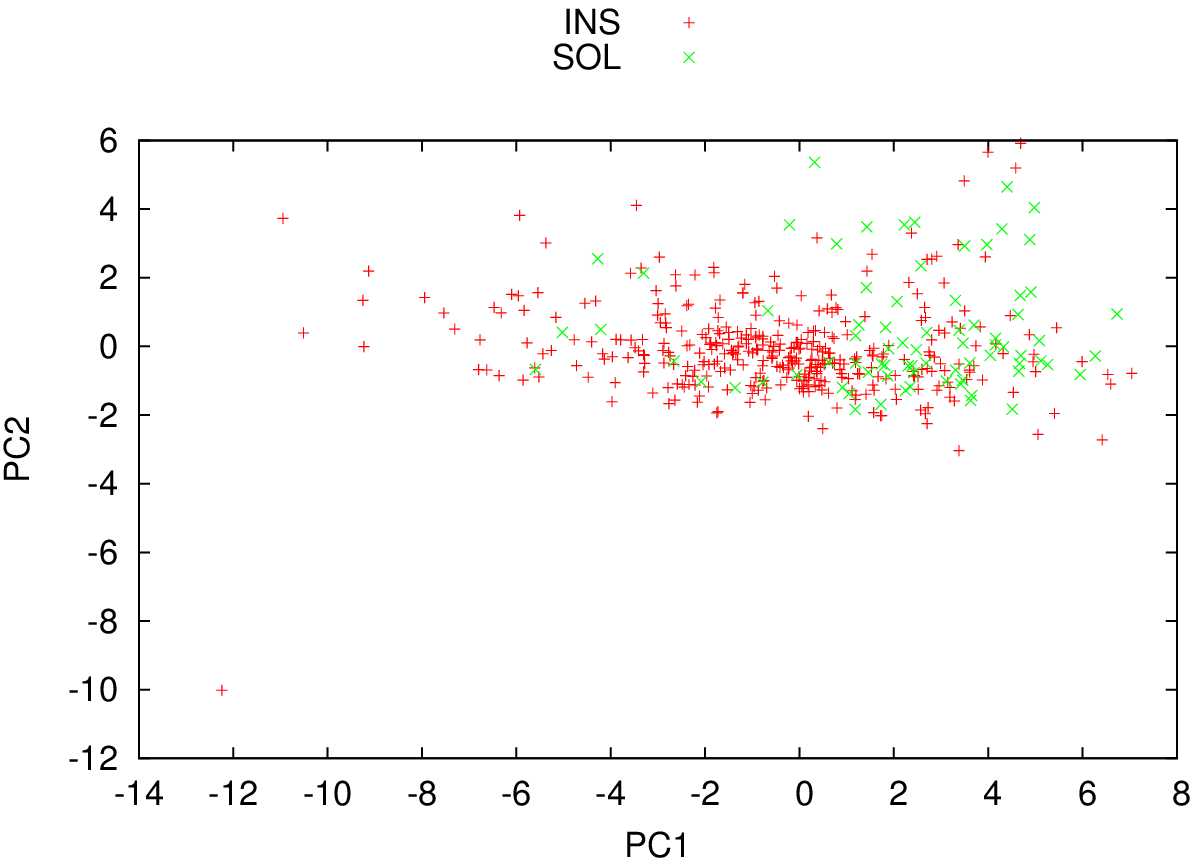}
\label{fig:pca_1-2}}
~
\subfigure[PC1--PC3.]{
\includegraphics[bb=0 0 343 245,scale=0.65,keepaspectratio=true]{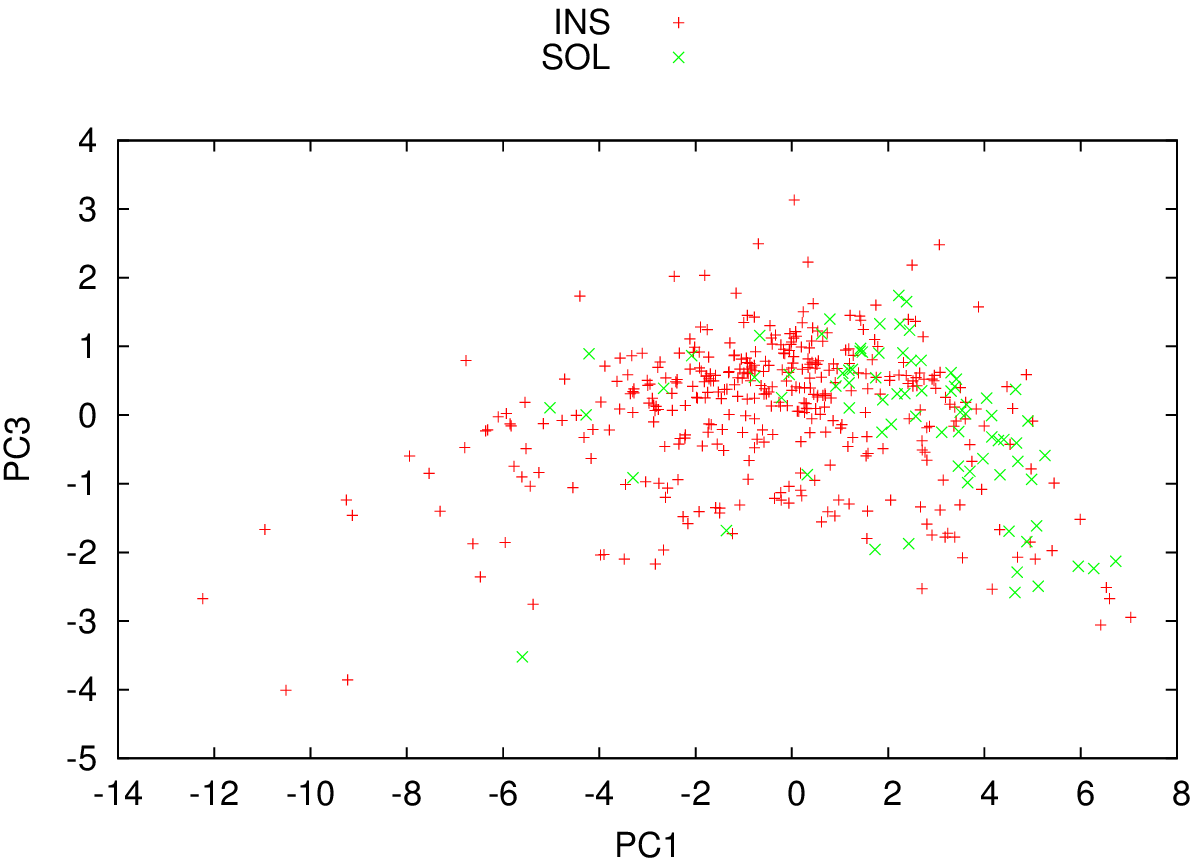}
\label{fig:pca_1-3}}

\caption{PCA of DS-C-454. First and second PCs (\ref{fig:pca_1-2}); first and third PCs (\ref{fig:pca_1-3}).}
\label{fig:pca_analysis}
\end{figure*}

Fig. \ref{fig:pca_analysis} shows the PCA of the data in DS-C-454.
In (\ref{fig:pca_1-2}) and (\ref{fig:pca_1-3}) we show, respectively, the plot of first--second and first--third PCs.
Data in DS-C-454 has been normalized by considering the component-wise mean a standard deviation.
As already demonstrated in Sec. \ref{sec:factors} by the factor analysis, PCA of the first two components, PC1--PC2, clearly shows some possibility of separating the patterns by the first component (i.e., the ``size'' of proteins).
According with the data in Tab. \ref{tab:factors}, in the experiments we will consider the original features transformed as the first five components obtained from the PCA of DS-C-454.

Split of disjoint training, validation, and test sets has been performed as follows. For the training set, we randomly select 50 soluble proteins. The remaining 27 are used for validating and testing the model, by using respectively 7 and 20 patterns.
The 377 insoluble proteins are considered in the validation and test sets, by using, respectively, 50 and 327 proteins.

In the case of soft decisions, EOCC outputs the membership degree (a number in $[0, 1]$) of a test pattern to the target class.
In fact, the model of EOCC consists in different decision regions (DRs) modeled as fuzzy sets \cite{eocc}.
The performance measure that we consider in this case is the Area Under the ROC Curve \cite{Fawcett:2006:IRA:1159473.1159475} (AUC), which can be interpreted as the average probability of ranking a target pattern higher than a non-target one.
For the hard decision case, we consider the confusion matrix and related common statistics: accuracy, precision, recall, F-measure, and false positive rate.
Finally, we provide an indicator of model complexity, which in our case is reported as the average number of DRs synthesized by EOCC.
Results are intended as the average of 30 different runs with different initialization seeds (we report standard deviations).

\subsubsection{Classification Results}

Test set results of EOCC are reported in Tab. \ref{tab:results}, considering both hard and soft decisions of membership to the target class (i.e., the class of very soluble proteins).
As the reader would notice, there are two rows in Tab. \ref{tab:results}: the first one referring to a ``Normal'' and the other to a ``Randomized'' scenario.
The one indicated as ``Normal'' provides the results considering exactly the split configuration as discussed above.
The row labeled ``Randomized'' is added to demonstrate the robustness of the results obtained in the ``Normal'' scenario.

Let us start by first discussing the results in the normal case. Average AUC is 0.74, denoting a fairly robust statistics when considering the soft decisions. Results from the confusion matrix (hard decisions) are less appealing. In fact, although the accuracy is fairly high (explained by the very low false positive rate), precision and recall are definitely low.
In average, EOCC synthesized 28.05 DRs, which divided by the training set size (i.e., 50) gives a compression factor of 0.44 (almost a DR every two training patterns). Such a fairly high number of DRs can be explained by the high difficulty of the underlying classification problem.
Considering the hard decisions, results are not convincing for what concerns the recognition of target patterns (they recall some of those obtained in our previous studies \cite{ecoli_graph,grapsec_ijcnn_2013}).

Let us focus now on the randomized scenario. In this case, we randomly defined 77 out of all 454 proteins as the target patterns, i.e., without considering their true association with respect to the soluble/insoluble class.
The remaining 377 patterns are accordingly labeled as non-target. We preserved the same split percentages for training, validation, and test, as those explained before. EOCC is then executed on such a randomized target/non-target assignment with the aim of evaluating the robustness of the results obtained in the normal case.
The second row of Tab. \ref{tab:results} shows the obtained results. Not surprisingly, the results for the hard classification are almost the same of those obtained in the normal setting. This is explainable by the fact that the problem is highly unbalanced (there are more non-target instances that affect the statistics derived from the confusion matrix).
However, it is possible to note two very important facts: (i) the AUC in this case is consistent with a random classifier (i.e., 0.5) and the average model complexity is now much higher.
In fact, in average 40.10 DRs are used, denoting a compression factor of roughly 0.19, which is considerably lower than the one obtained in the normal case (2.3 times lower).
This is a first indicator that the solution obtained in the normal case can be considered as robust, i.e., a model that effectively provides a reasonable explanation of the underlying process.

\begin{figure*}[ht!]
\centering

\subfigure[]{
\includegraphics[bb=0 0 340 243,scale=0.65,keepaspectratio=true]{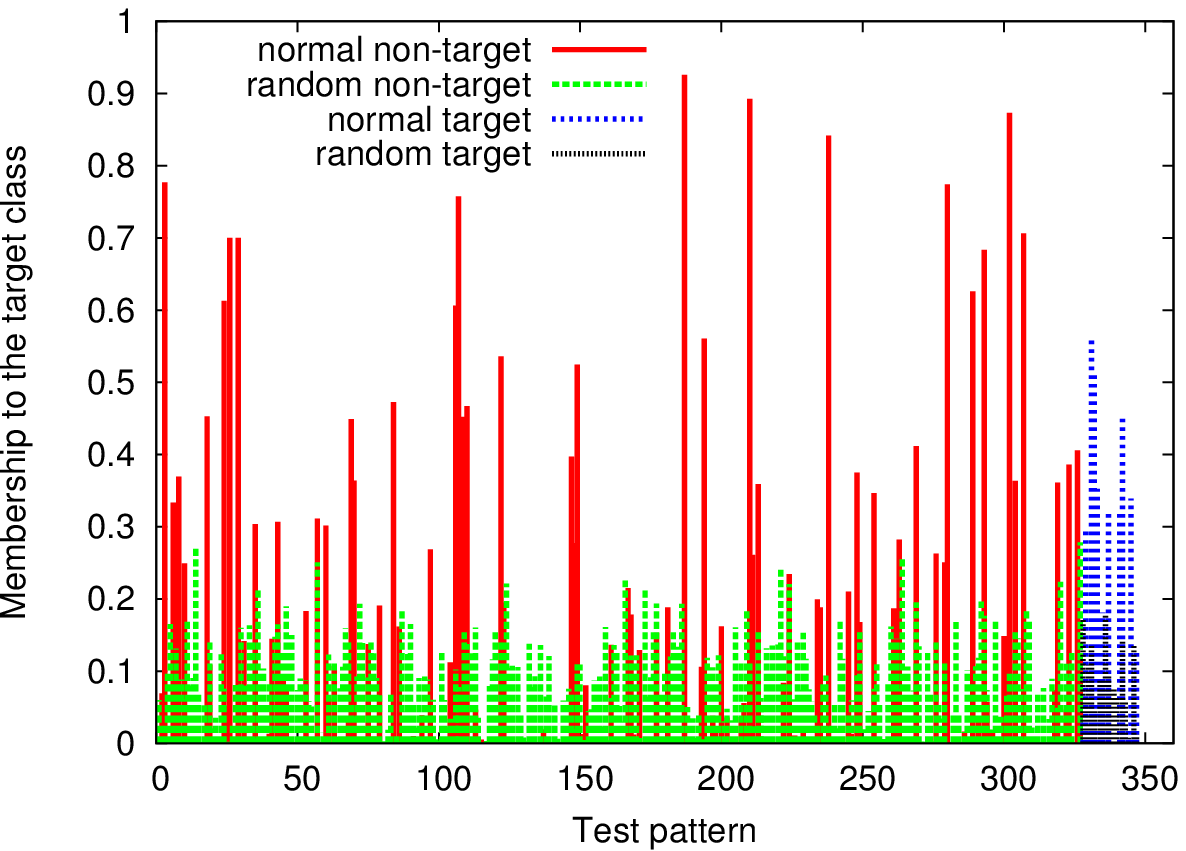}
\label{fig:memb}}
~
\subfigure[]{
\includegraphics[bb=0 0 340 243,scale=0.65,keepaspectratio=true]{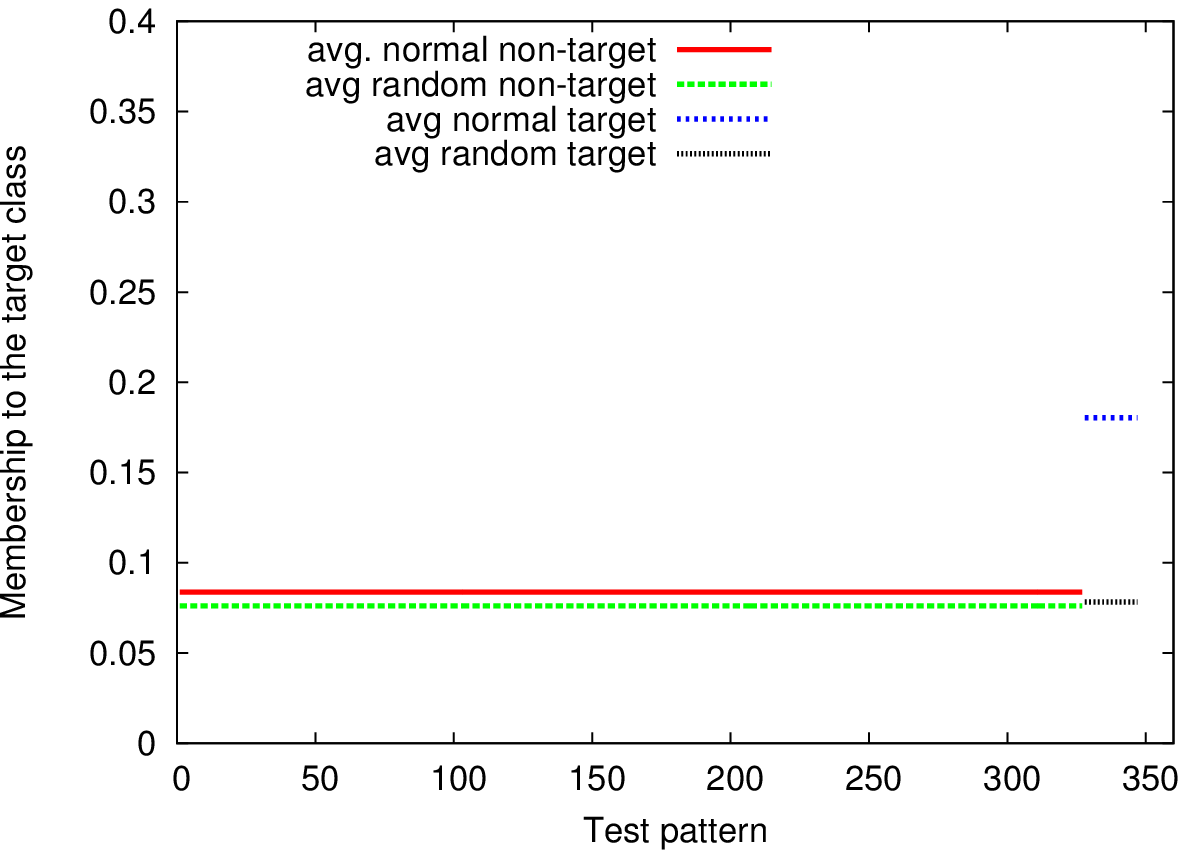}
\label{fig:avg}}

\caption{Membership degrees to the target class assigned to the test patterns. In Fig. \ref{fig:memb} we show the memberships of target and non-target patterns, for either the normal and randomized target class setting. Fig. \ref{fig:avg} reports the averages of those membership degrees.}
\label{fig:memberships}
\end{figure*}

As a second demonstration of this conclusion, let us analyze Fig. \ref{fig:memberships}.
In these two figures, we compare the continuous outputs calculated by EOCC in the two considered scenarios.
In Fig. \ref{fig:memb} we report the membership degrees assigned to the test patterns, while Fig. \ref{fig:avg} reports the average of those values. As it is possible to note, in the figures we differentiate among those of the normal and randomized setting, and also among those patterns that are considered as target and non-target.
From these results it is possible to deduce that the membership degrees of the target patterns are highly affected by the randomization of the target class, while those of the non-target patterns are practically left invariant.
However, membership values, in the normal case, denote a good average class discrimination, i.e., the tested target patterns have an average membership degree fairly higher than the non-target instances (although it is possible to note different errors for the non-target patterns). This is in in clear accordance with the fact that, originally, proteins are characterized by a continuous solubility degree.
To conclude, it is worth stressing that EOCC is a classifier, and so the soft decisions must be intended as a way to assign a score (ranking) to the decisions, and not as an approximation of the original solubility signal.
\begin{table*}[thp!]\scriptsize
\begin{center}
\caption{Test set results considering both soft and hard decisions of membership to the target class.}
\label{tab:results}
\begin{tabular}{|c|c|c|c|c|c|c|c|}
\hline
 & \textbf{\textit{Soft Decision}} & \multicolumn{5}{c|}{\textbf{\textit{Hard Decision}}} & \\
\hline
 & \textbf{AUC} & \textbf{Accuracy} & \textbf{Precision} & \textbf{Recall} & \textbf{F-Measure} & \textbf{False Positive Rate} & \textbf{\# DRs} \\
\hline
Normal & 0.74(0.02) & 0.91(0.02) & 0.12(0.10) & 0.11(0.09) & 0.10(0.08) & 0.04(0.03) & 28.05(4.15) \\
\hline
Randomized & 0.50(0.06) & 0.91(0.03) & 0.05(0.07) & 0.02(0.05) & 0.03(0.04) & 0.03(0.03) & 40.10(6.28) \\
\hline
\end{tabular}
\end{center}
\end{table*}

\section{Conclusions and Future Directions}
\label{sec:conclusions}

The relation among the structure and solubility property of proteins is highly non-linear and hence very hard to predict.
However, solubility/aggregation propensity of proteins is a very important topic, which justifies such a research effort.
In fact, aggregation of proteins is at the basis of many misfolding diseases, such as Parkinson and Alzheimer.

The herein presented study analyzed three important factors of a dataset of proteins recently elaborated from the E.Coli proteome. First, we checked pure topological characteristics, by analyzing properties related to the shape (fractal dimension of the embedded graphs) and connections (distribution of shortest paths and vertex degrees). We verified that also this dataset of E.Coli proteins adhere to general principles describing proteins architecture.
Then we moved to the statistical analysis of a collection of 15 features that we extracted from the dataset of graphs under analysis.
We studied either linear and non-linear relations among such features. Overall, our findings confirmed that the solubility of proteins is mainly distinguished by their size. In fact, the size affects the density of the connections, the global shape, and thus necessarily also the propensity of being soluble. This last fact is due to evident energetic constraints, which favors the (completely autonomous) folding of small-sized proteins.
Lastly, we analyzed the proteins described in terms of those 15 features, for the purpose of recognizing very soluble and poorly soluble proteins automatically. We faced the problem in the one-class classification setting by means of a novel method (EOCC).
In the case of hard discrimination, results are not convincing. However, results concerning the soft decisions yielded interesting performances in terms of discrimination of the solubility class.

Notwithstanding we reached encouraging recognition performances (in the soft decision case), an ultimate solution in this sense is still missing.
The extreme difficulty of this problem may be due to a variety of reasons: (i) a highly non-linear relation among structure and solubility, (ii) experimental errors in the measurements of the original solubility degree, (iii) a too small sample that is not sufficient to synthesize an effective recognition model, and/or (iv) a wrong modeling and computational approach.
All in all, we were aware of the intrinsic difficulty of the problem, given solubility and aggregation can be considered as the ``two-faces of the same coin'' (being the coin the ability to form non-covalent bonds) more than a neat categorization into two non-overlapping behaviors.
We want to stress how the quest for solving a difficult problem allowed us to find along the way a novel style of graph-based data analysis approach that, in the case of proteins, allowed us to highlight some very important and general properties of protein structure and function.
Protein molecules are an almost unique case in which graphs have an immediate physical counterpart, but the same approach can be equally applied to more abstract / less material network structures, such as gene expression \cite{pinna2011simulating} or metabolic \cite{tun+dhar+palumbo+giuliani2006} networks.

\appendices

\section{Graph Characteristics}
\label{sec:graph_descriptors}

\subsection{Radius of Gyration and Porosity}

Radius of gyration of a molecule (graph) of $N$ amino acids (vertices) is computed as \cite{di2012proteins},
\begin{equation}
R_{G} = \sqrt{\frac{\sum_{i=1}^{N} m_i r_{i}^{2}}{2\sum_{i=1}^{N} m_i}},
\end{equation}
where $m_i$ is the mass of the \textit{i}th amino acid and $r_i$ is the Euclidean distance between the amino acid center of mass and the center of mass of the whole structure.

The porosity (or void fraction) of a molecule is defined as
\begin{equation}
\mathrm{P}=1 - \frac{V_{\mathrm{residue}}}{V},
\end{equation}
where $V_{\mathrm{residue}}$ is the sum of the volumes of the residues constituting the molecule.
$V$ is the total volume, which is computed as the average of the three spherical volumes $V_x, V_y, V_z$, each one calculated by considering a diameter equal to the maximum distance in the respective dimension.

\subsection{Modularity of a Graph}

A partition \cite{Livi_ga_2013}, $K(G)$, of order $k$ of a graph $G=(\mathcal{V}, \mathcal{E})$, is commonly intended as a partition of the vertex set $\mathcal{V}(G)$ into disjoint subsets (clusters, modules), $K(G)=\{\mathcal{C}_1, \mathcal{C}_2, ..., \mathcal{C}_k\}$.
A well-established measure to determine the quality of $K(G)$ is the so-called modularity measure \cite{fortunato2010}, which basically quantifies how well $K(G)$ groups the vertices of $G$ into compact and separated clusters.
Intuitively, in a graph a cluster of vertices is compact if the number (the weight) of the intra-cluster edges is considerably greater than the one of the inter-cluster edges.
The modularity measure is formally defined as follows,
\begin{align}
\label{eq:modularity}
M(G, K(G)) = \frac{1}{2|\mathcal{E}(G)|}\sum_{i=1}^{k}\sum_{j=1}^{k} \left( A_{ij}-\frac{\mathrm{deg}(v_i)\mathrm{deg}(v_j)}{2|\mathcal{E}(G)|} \right) \delta(\mathcal{C}_i, \mathcal{C}_j),
\end{align}
where $\mathrm{deg}(v_i)$ is the number of edges incident to $v_i$.
Eq. \ref{eq:modularity} can be rewritten as:
\begin{equation}
\label{eq:modularity2}
M(G, K(G)) = \sum_{l=1}^{k} \left[ \frac{|\mathcal{E}(\mathcal{C}_l)|}{|\mathcal{E}(G)|}-\left( \frac{\mathrm{deg}(\mathcal{C}_l)}{2|\mathcal{E}(G)|} \right)^2 \right].
\end{equation}

$|\mathcal{E}(\mathcal{C}_l)|$ is the number of intra-cluster edges and $\mathrm{deg}(\mathcal{C}_l)$ is the sum of degrees of the vertices in the l\textit{th} cluster (considering all edges, i.e., also those with one end-point outside $\mathcal{C}_i$).
The modularity of a graph $G$ is equal to the modularity of the partition of $G$ that maximizes Eq. \ref{eq:modularity2}.
Finding such an optimal partition (\ref{eq:modularity2}) is NP-complete \cite{4358966}, and therefore many heuristics has been proposed \cite{fortunato2010}.

\subsection{Closeness and Degree Centrality}

In a graph $G=(\mathcal{V}, \mathcal{E})$, the closeness centrality factor of a vertex $v\in\mathcal{V}$ is defined as:
\begin{equation}
\mathrm{CC}(v)=\sum_{u\in\mathcal{V}, u\neq v} 2^{-d(v, u)},
\end{equation}
where $d(\cdot, \cdot)$ computes the shortest path in $G$. The closeness centrality provides a measure of how much a vertex is close to all other vertices of the graph (in terms of shortest paths). Such measure can be extended to the whole graph by taking the average among all vertices.

The degree centrality of a vertex $v\in\mathcal{V}$ is defined as its degree, i.e., the number of incident edges. Analogously, it can be normalized by using the maximum degree of the graph.

\subsection{Clustering Coefficient}

Let $\mathbf{A}$ be the adjacency matrix of a graph $G$. 
The clustering coefficient \cite{costa2007characterization} is defined as,
\begin{equation}
\mathrm{CL}=\frac{3N_{\Delta}}{N_{3}},
\end{equation}
where $N_{\Delta}$ is the number of triangles in the graph, while $N_3$ is the number of connected triplets:
\begin{align}
N_{\Delta} =&\sum_{k>j>i} A_{ij} A_{ik} A_{jk}; \\
N_3 =& \sum_{k>j>i} (A_{ij}A_{ik} + A_{ji}A_{jk} + A_{ki}A_{jk}).
\end{align}

\subsection{Energy and Laplacian Energy of a Graph}

Let $G=(\mathcal{V}, \mathcal{E}), n=|\mathcal{V}|, m=|\mathcal{E}|$, be a graph, and let $\mathbf{A}^{n\times n}$ be its adjacency matrix.
The energy \cite{gutman2006laplacian} of the graph is defined as
\begin{equation}
\mathrm{E} = \sum_{i=1}^{n} |\lambda_{i}|,
\end{equation}
where $\lambda_{i}$ is the \textit{i}th eigenvalue of the adjacency matrix $\mathbf{A}$.

Let us define the Laplacian matrix as $\mathbf{L}=\mathbf{D} - \mathbf{A}$, where $\mathbf{D}$ is a diagonal matrix containing the vertex degrees.
The Laplacian energy \cite{gutman2006laplacian}, LE, is defined as
\begin{equation}
\mathrm{LE} = \sum_{i=1}^{n} \left|\mu_{i} - \frac{2m}{n}\right|,
\end{equation}
where $\mu_{i}$ is the \textit{i}th eigenvalue of the Laplacian.

\subsection{Heat Kernel}

Let $G=(\mathcal{V}, \mathcal{E})$ be a graph with $n$ vertices and $m$ edges, respectively.
Let $\mathbf{A}$ and $\mathbf{L}$ be the corresponding adjacency and Laplacian matrices.
Let us define the normalized Laplacian matrix as $\hat{\mathbf{L}}=\mathbf{D}^{-1/2}\mathbf{L}\mathbf{D}^{-1/2}$.
$\hat{\mathbf{L}}$ is symmetric and positive semi-definite, and therefore it has non-negative eigenvalues only.
The spectral decomposition of the Laplacian is given by $\hat{\mathbf{L}}=\Phi \Lambda \Phi^{T}$, where $\Lambda$ is the diagonal matrix containing the eigenvalues arranged as $0=\lambda_1, \lambda_2, ..., \lambda_n$; $\Phi$ contains the corresponding (unitary) eigenvectors.

The heat equation \cite{Xiao:2009:GCH:1563046.1563099,suau2013analysis,kondor02Diffusionkernelsgraphsotherdiscretestructures} associated to the Laplacian $\hat{\mathbf{L}}$ is given by
\begin{equation}
\label{eq:heat_equation}
\frac{\partial h_{t}}{\partial t} = -\hat{\mathbf{L}}h_{t},
\end{equation}
where $h_{t}$ is the heat kernel (a doubly-stochastic $n\times n$ matrix) and $t$ is the time variable.
It is well-known that the solution for (\ref{eq:heat_equation}) is,
\begin{equation}
h_{t} = \exp(-t\hat{\mathbf{L}}),
\end{equation}
which can be solved by exponentiating the spectrum of $\hat{\mathbf{L}}$:
\begin{equation}
h_{t} = \Phi \exp(-\Lambda t) \Phi^{T} = \sum_{i=1}^{n} \exp(-\lambda_i t)\phi_i\phi_i^{T}.
\end{equation}

Eq. \ref{eq:heat_equation} describes the diffusion (i.e., the flow) of heat/information across the graph over time. In fact,
\begin{equation}
h_{t}(v, u) = \sum_{i=1}^{n} \exp(-\lambda_i t)\phi_i(v)\phi_i(u),
\end{equation}
where $\phi_i(v)$ denotes the value related to $v$ in the \textit{i}th eigenvector.
It is important to note (\cite{Xiao:2009:GCH:1563046.1563099}) that $h_{t} \simeq \mathbf{I} - \hat{\mathbf{L}}t$ when $t\rightarrow 0$; conversely, when $t$ is large we have $h_{t}\simeq \exp(-\lambda_2 t)\phi_2\phi_{2}^{T}$ (note that $\phi_2$, i.e., the eigenvector associated to smallest non-zero eigenvalue, is called Fiedler vector).
This means that the large-time behavior of the diffusion depends on the global structure of the graph, while its short-time characteristics are determined by the local connections.

The heat trace (HT) of $h_{t}$ is the sum of the diagonal entries,
\begin{equation}
\label{eq:heat_trace}
\mathrm{HT} = \mathrm{Tr}(h_{t}) = \sum_{i=1}^{n} \exp(-\lambda_i t),
\end{equation}
which takes into account only the eigenvalues of $\hat{\mathbf{L}}$.
The heat content of $h_t$ is defined by considering also the eigenvectors:
\begin{align}
\label{eq:heat_content}
Q(t) = \sum_{u\in\mathcal{V}} \sum_{u\in\mathcal{V}} h_{t}(u, v) = \sum_{u\in\mathcal{V}} \sum_{u\in\mathcal{V}} \sum_{i=1}^{n} \exp(-\lambda_i t)\phi_i(v)\phi_i(u).
\end{align}

Eq. \ref{eq:heat_content} can be described in terms of power series,
\begin{equation}
\label{eq:heat_coeff}
Q(t) = \sum_{m=0}^{\infty} q_m t^m.
\end{equation}

The McLaurin series expansion for the negative exponential reads as:
\begin{equation}
\exp(-\lambda_i t)=\sum_{m=0}^{\infty} \frac{(-\lambda_i)^{m} t^m}{m!},
\end{equation}
which substituted in Eq. \ref{eq:heat_content} gives:
\begin{align}
Q(t) = \sum_{u\in\mathcal{V}} \sum_{u\in\mathcal{V}} \sum_{i=1}^{n} \exp(-\lambda_i t)\phi_i(v)\phi_i(u) = \sum_{m=0}^{\infty}\sum_{u\in\mathcal{V}} \sum_{u\in\mathcal{V}} \sum_{i=1}^{n} \phi_i(v)\phi_i(u) \frac{(-\lambda_i)^{m} t^m}{m!}.
\end{align}

The coefficients $q_m$ in (\ref{eq:heat_coeff}) are graph invariants (called heat content invariants, HCI) that be obtained in closed-form via the following expression:
\begin{equation}
q_m = \sum_{i=1}^{n} \left\{ \left( \sum_{u\in\mathcal{V}} \phi_{i}(u) \right)^2 \right\} \frac{(-\lambda_i)^m}{m!}.
\end{equation}

\subsection{Ambiguity of a Graph}

The ambiguity of a graph $G=(\mathcal{V}, \mathcal{E}), n=|\mathcal{V}|$, gives a measure of uncertainty elaborated according to a fuzzy set based interpretation \cite{Livi_ga_2013}.
The ambiguity of the graph $G$ is calculated by embedding the graph into a fuzzy hypercube $\mathcal{I}=[0, 1]^n$, which, in short, encodes the membership values of the vertices.
A graph $G$ is mapped to a type-1 fuzzy set $\mathcal{F}$, defined as
\begin{equation}
\mathcal{F}=\{(v, \mu_{\mathcal{F}}(v)) | \ v\in\mathcal{V}(G), \mu_{\mathcal{F}}(v)\in[0, 1] \},
\end{equation}
by generating the membership function $\mu_{\mathcal{F}}(\cdot)$ of the graph vertices, $\mathcal{V}(G)$.
Such a membership function is constructed by considering a partition, $P$, of the graph:
\begin{equation}
P=\bigcup_{i=1}^{k} \mathcal{C}_i.
\end{equation}

The partition $P$ is then ``fuzzified'' by computing the t-conorm $\bot$ among all fuzzy sets $\mathcal{F}_i$ associated to each $\mathcal{C}_i$ (one-to-one mapping), yielding the resulting fuzzy set, $\mathcal{F}$, describing the graph as a whole:
\begin{equation}
\label{eq:fuzz}
\mathcal{F}= \displaystyle\bot_{i=1}^{k} \mathcal{F}_i.
\end{equation}

In the following, let us write $\mathcal{F}=\phi^{P}(G)$.
The membership function $\mu_{\mathcal{F}_{i}}(\cdot)$ describing the fuzzy set $\mathcal{F}_i$ is generated according to the following expression:
\begin{equation}
\mu_{\mathcal{F}_{i}}(v) = \alpha_{\mathcal{C}_{i}}(v) \times \beta_{\mathcal{C}_{i}}(v).
\end{equation}

$\alpha_{\mathcal{C}_{i}}(v)$ accounts for the degree concentration $v$ in $\mathcal{C}_i$, while $\beta_{\mathcal{C}_{i}}(v)$ gives the importance of the vertex in $\mathcal{C}_i$ in terms of centrality.
Given the fuzzy set $\mathcal{F}$ representing the uncertainty of the whole graph $G$, the measure of ambiguity of $G$, denoted with $A(G)$, is obtained by computing (any monotonic and non-decreasing transformation of) the fuzzy entropy of $\mathcal{F}$.
Since there are exponentially-many fuzzy set representations for a single graph $G$, the actual ambiguity value is calculated as the solution of the following combinatorial optimization problem:
\begin{equation}
A(G) = \min_{P} A(\phi^{P}(G)).
\end{equation}

$A(G)$ assumes values within the $[0, 1]$ range, approaching one as the graph is maximally ambiguous (i.e., maximally irregular). It has been proved that $A(G)$ is zero when the graph is regular (e.g., complete) \cite{Livi_ga_2013}. Accordingly, $A(G)$ can be used as a global complexity descriptor characterizing the regularity of $G$.

\subsection{Entropy of a Markovian Random Walk}

A Markovian random walk in a graph $G=(\mathcal{V}, \mathcal{E})$ \cite{Dehmer201157} is a first-order Markov chain that generates sequences of vertices (the vertices of the graph should be interpreted as the states of chain).
Transition among vertices are regulated by the transition matrix, which is given by $\mathbf{T}=\mathbf{D}^{-1}\mathbf{A}$.
Let $\underline{\mathbf{p}}_{t}$ be a probability vector describing the probability of the states at time $t$; when $t=0$, the vector describes the initial distribution.
The stationary distribution is a probability vector, $\pi$, such that $\pi\mathbf{T}=\pi$. The stationary distribution of a random walk can be intended as the limiting distribution, i.e., the distribution of the vertices/states when $t\rightarrow\infty$.
If the graph is undirected and non-bipartite, then the random walk always admits a stationary distribution, which can be easily computed from the degree distribution:
\begin{equation}
\pi_i = \frac{\mathrm{deg}(v_i)}{2|\mathcal{E}|}.
\end{equation}

A stationary random walk is hence completely described by $\pi$. The entropy of $\pi$ can be used to characterize $G$ in terms of predictability of the corresponding stationary random walk. In fact, if $G$ is regular, then $\pi$ is uniform, which could be interpreted as the maximum degree of unpredictability.

\bibliographystyle{abbrvnat}
\bibliography{/home/lorenzo/University/Research/Publications/Bibliography.bib}
\end{document}